\documentclass[acmsmall,screen,nonacm]{acmart}

\usepackage{xspace}
\usepackage{enumitem}
\usepackage{multirow}
\usepackage{bbding}
\usepackage{cancel}

\usepackage{array}
\usepackage{booktabs}
\usepackage{tabularx}
\usepackage{colortbl}
\usepackage{soul}

\newcolumntype{Y}{>{\raggedright\arraybackslash}X}

\newcommand\tool{\textsc{Req2Test-Bench}\xspace}

\usepackage{tikz}
\newcommand{\stepcircle}[1]{%
    \tikz[baseline=-0.7ex]{
    \node[circle, fill=black, text=white, font=\sffamily\scriptsize\bfseries,
      inner sep=0.35pt, minimum size=0.75em] (char) {#1};
  }%
}

\AtBeginDocument{%
  }

\begin{document}

\title{Benchmarking LLM Agents for Behaviorally Adequate Test Generation from Software Requirements}

\author{Binhang Qi}
\email{qibh@nus.edu.sg}
\affiliation{%
  \institution{National University of Singapore}
  \country{Singapore}}

\author{Yiming Liu}
\email{liu\_yiming@sjtu.edu.cn}
\affiliation{%
  \institution{Shanghai Jiao Tong University}
  \country{China}}
\affiliation{%
  \institution{Shanghai Innovation Institute}
  \country{China}}

\author{Yun Lin}
\authornote{Corresponding Author.}
\email{lin\_yun@sjtu.edu.cn}
\affiliation{%
  \institution{Shanghai Jiao Tong University}
  \country{China}}

\author{Yangjie Zhou}
\email{yj\_zhou@nus.edu.sg}
\affiliation{%
  \institution{National University of Singapore}
  \country{Singapore}}

\author{Hailong Sun}
\email{sunhl@buaa.edu.cn}
\affiliation{%
  \institution{Beihang University}
  \country{China}}

\author{Jin Song Dong}
\email{dcsdjs@nus.edu.sg}
\affiliation{%
  \institution{National University of Singapore}
  \country{Singapore}}

\begin{abstract}
As large language model (LLM) agents increasingly compile natural-language requirements into repository-level code, requirement-to-test generation grows in importance to provide validation constraints (or harness) to guide code generation.
Yet it remains unclear whether current agents can generate tests that validate the required behavior when the implementation of that behavior is withheld.
Existing test-generation benchmarks mostly focus on white-box testing, whereas the requirement--test-related study usually requires evaluation on semantic alignment between the given documents and the generated tests.

We introduce \tool, a benchmark containing 5,086 execution-validated instances recovered from developer-authored behavior-driven development (BDD) artifacts in 8 open-source projects, spanning 5 programming languages and 4 BDD frameworks.
Each instance pairs a natural-language requirement in Gherkin with its developer-written \textit{Test Implementation}.
Given the requirement, target interfaces of the required behavior, and a repository view that withholds either the target implementation or all production implementations, an agent must generate a complete, potentially multi-file \textit{Test Implementation}.
Solving the task requires bridging the gap between a high-level requirement and its test-code correspondence regarding project-specific knowledge: \textit{crystallizing \ul{what to test}, stated in natural language, into \ul{how to test it} with the project's own fixtures, helpers, and interfaces.}
We evaluate the semantic difference between generated and ground-truth tests by comparing their production API calls, code coverage, and mutant kills.
Evaluation of 8 agent--model--reasoning configurations reveals a substantial gap between execution success and behavioral adequacy, which is measured strictly as kill-completeness: a kill-complete test kills every mutant that the ground-truth test kills.
The best-performing configuration, Codex with DeepSeek-V4.1-Flash, produces passing tests for 82.8\% of instances, yet only 33.8\% are behaviorally adequate.
We find that 
(1) scaling up agents yields sharply diminishing returns in behavioral adequacy relative to generation cost;
(2) under-specified requirements cause the information gap, including ambiguous setup and implicit expected outcomes; and
(3) requirement-grounded harnesses that steer agent non-determinism offer a promising direction for improvement.

\end{abstract}

\maketitle

\section{Introduction}
Large language model (LLM) agents increasingly translate natural-language software requirements into repository-wide implementations~\cite{wang2025openhandsopenplatformai}.
As code generation accelerates, validation becomes a bottleneck:
developers need executable evidence that generated software satisfies the requested behavior rather than merely compiling or appearing plausible.
Requirement-derived tests provide behavioral checks for code-generation
harnesses, supporting candidate selection and iterative
refinement~\cite{chen2022codet,huang2023agentcoder}. 
Yet a generated test may pass while omitting meaningful checks or relying
on tautological assertions~\cite{schafer2023empirical,jain2025testgeneval}.
The challenge is sharper when tests must be written before an implementation of the required behavior is available~\cite{kong2026arc}.
We thus ask: \textit{\textbf{can current LLM agents generate complete, executable, and behaviorally adequate tests from software requirements when the target implementation is withheld?}}
We examine it through three questions:
\begin{itemize}[leftmargin=*]
    \item Q1: Does scaling up LLM agents reliably and proportionally improve behavioral adequacy?
    \item Q2: If not, what gaps limit the progress?
    \item Q3: What directions are promising to narrow the gap, guiding the development of LLM agents?
\end{itemize}

In this work, we define the requirement and the behavioral adequacy as follows:

\vspace{3pt}

\noindent\textsc{Definition} 1. (\textbf{Requirement})
\textit{A requirement is a natural-language statement of one software behavior expected from a software system, stated independently of the code that realizes it.
It is concrete enough to determine a testing objective, typically by stating the relevant conditions, the triggering interaction, and the expected observable outcome.}

\vspace{3pt}

\noindent\textsc{Definition} 2 
(\textbf{Behavioral Adequacy}).
\textit{A test is behaviorally adequate for a requirement if it exercises the behavior prescribed by that requirement and correctly checks the expected outcomes in the project's context.}

\vspace{3pt}

Existing LLM-based test-generation studies and benchmarks start from prescribed methods, classes, or source files~\cite{tufano2022methods2test,zhang2024testbench,jain2025testgeneval}, from code changes~\cite{hu2026difftestgen}, from issue reports~\cite{mundler2024swt,ahmed2024tddbench}, or from behavior inferred by exploring a running application~\cite{alian2025feature}.
Consequently, their results alone do not suffice to answer these questions.

\begin{figure}[t]
    \centering
    \includegraphics[width=\linewidth]{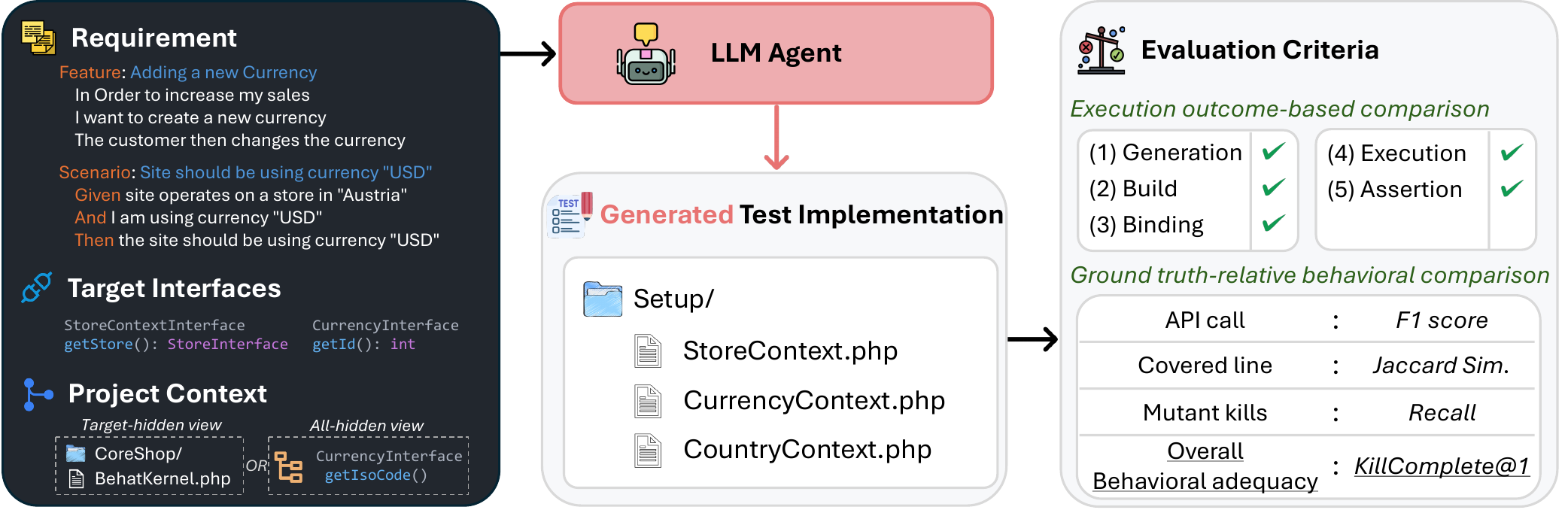}
    \caption{Overview of the \tool task. 
    Each instance provides a \textit{requirement} composed of enclosing \texttt{Feature} context and a target \texttt{Scenario}, together with designated \textit{target interfaces} and \textit{project context} exposed through either a \textit{target-hidden} or an \textit{all-hidden} view.
    Both views withhold the target implementation.
    Target-hidden retains other permitted production implementations, whereas all-hidden withholds them.
    An LLM agent generates a complete, potentially multi-file \textit{Test
    Implementation}.
    After generation, an evaluator runs the test against the complete production implementation, recording staged functional outcomes and comparing its production-API calls, code coverage, and mutant kills with those of the ground-truth Test Implementation.}
    \label{fig:req2test}
    \vspace{-10pt}
\end{figure}

We instead study requirement-to-test generation, which is \emph{requirement-driven}: an
independently stated requirement determines the testing objective, and the target implementation is withheld from the agent.
Related requirement--test resources address parts of this problem; however, they have complementary limitations:
\stepcircle{1}~Some synthesize or manually adapt requirement or test artifacts during benchmark construction~\cite{liu2026e2edev,li2024devbench};
\stepcircle{2}~others associate a coarse requirement artifact, such as a project-level requirement, with an entire test file rather than with the complete Test Implementation for one individual, concrete behavior~\cite{
li2024devbench,junior2026givenwhenthen}.

To address these limitations, \tool combines authentic requirements, complete ground-truth Test Implementations for individual behaviors, execution-validated mappings, and behavioral evaluation under controlled implementation visibility.
We instantiate the task using artifacts from behavior-driven-development (BDD) repositories.
BDD projects express requirements in Gherkin, where a \texttt{Feature} describes a capability and groups related \texttt{Scenario}s, while a \texttt{Scenario} expresses a concrete behavioral example through an ordered sequence of \texttt{Step}s~\cite{cucumberGherkin}, as shown in \autoref{fig:req2test}.
Their availability across open-source projects and programming languages, without restriction to a particular test level, supports large-scale and diverse benchmark construction~\cite{junior2026givenwhenthen}.

Accordingly, for each \tool instance, we operationally define one requirement $r_i$ as the pair of an enclosing \texttt{Feature} $F_i$ and a target \texttt{Scenario} $S_i$.
We formalize the requirement-to-test generation task as:
\begin{equation}
    T_i^{gen}
      =
    \mathcal{A}\!\left(r_i,\, \mathcal{I}^{\alpha}_i\right),
    \qquad
    r_i \triangleq \langle F_i,S_i\rangle,
    \qquad
    \mathcal{I}^{\alpha}_i \triangleq \bigl\langle \mathcal{I}^{\mathrm{tar}}_i,\, C_{\alpha}(P_i)\bigr\rangle .
    \label{eq:req2test}
\end{equation}
The agent-visible project information $\mathcal{I}^{\alpha}_i$ combines the designated target interfaces $\mathcal{I}^{\mathrm{tar}}_i$ with available project context $C_{\alpha}(P_i)$.
$\mathcal{I}^{\mathrm{tar}}_i$ provides method declarations derived from first-party public production calls made directly by the ground-truth test's action and check \texttt{Step}s.
They provide interface-localization cues, but do not enumerate the auxiliary APIs, fixtures, or helpers needed to construct the test.
$C_{\alpha}(P_i)$ of project snapshot $P_i$ is exposed through one of two views, with $\alpha\in\{\mathrm{TH},\mathrm{AH}\}$:
\stepcircle{1}~In the \emph{target-hidden view} $C_{\mathrm{TH}}(P_i)$, the selected production bodies realizing the target behavior, including relevant first-party delegation, are replaced with declaration-preserving placeholders.
\stepcircle{2}~In the \emph{all-hidden view} $C_{\mathrm{AH}}(P_i)$, implementation bodies throughout the project's declared production-code scope are likewise replaced.
The output $T_i^{gen}$ generated by an LLM agent $\mathcal{A}$ is one complete, potentially multi-file Test Implementation containing the test-side artifacts needed to build, bind, execute, and check $S_i$.
After generation, $T_i^{gen}$ from either view is evaluated by executing it against the same complete production implementation in a separate environment.
As illustrated by \autoref{fig:req2test}, $r_i$ pairs the \texttt{Feature} ``Adding a new Currency ...'' with the \texttt{Scenario} ``Site should be using currency `USD'\,...''. 
$\mathcal{I}^{\mathrm{tar}}_i$ includes \texttt{StoreContextInterface::getStore()} and \texttt{CurrencyInterface::getId()}, and $T_i^{gen}$ spans three PHP files.

To construct \tool, we recover the developer-written Test Implementation of each \texttt{Scenario} from the \texttt{Step Definition}s that the native BDD framework binds to its \texttt{Step}s at runtime, validate each recovered mapping by execution, and freeze a per-instance mutant pool with the kills of the ground-truth test.
\tool's \textit{validated mapping dataset} contains 5,086 Requirement--Test Implementation instances from 1,415 Features in 8 projects, spanning 5 languages, 6 domains, 4 BDD frameworks, and component-, integration-, and system-level tests.

We evaluate 8 agent--model--reasoning configurations that vary the coding agent (Aider, OpenHands, Claude Code, and Codex CLI), the LLM (DeepSeek-V4.1-Flash, GPT-6 Luna, GPT-6 Sol, and Claude Sonnet 5.5), and the reasoning effort.
Beyond execution success (Pass@1), we assess behavioral adequacy with the developers' own test as the reference for the required behavior.
Our primary metric, KillComplete@1, is the share of instances whose generated test both passes and kills every mutant that its ground-truth test kills, i.e., is \emph{kill-complete}.

The experimental results show that current agents produce passing tests far more often than behaviorally adequate ones.
The best-passing configuration, Codex CLI with DeepSeek-V4.1-Flash at high reasoning effort, passes 82.8\% of instances but is kill-complete on only 33.8\%.
Three findings emerge:
\stepcircle{1} \textbf{Returns diminish steeply}:
The returns in behavioral adequacy obtained by scaling up agents (e.g., increasing LLM scale) diminish with generation cost.
For Codex CLI, moving from GPT-6 Luna to DeepSeek-V4.1-Flash and then to GPT-6 Sol increases the generation cost by factors of 6.0 and 4.9, yet changes KillComplete@1 by only $+5.5$ percentage points (pp) and then $-2.5$~pp (28.3\% $\rightarrow$ 33.8\% $\rightarrow$ 31.3\%).
\stepcircle{2} \textbf{Information gap between high-level requirements and concrete tests}: 90--99\% of passing but kill-incomplete tests fall short where the requirement leaves the setup ambiguous or the expected outcome tacit.
\stepcircle{3} \textbf{A requirement-grounded harness is required for reducing uncertainty in test generation}: 
The best of three generations is kill-complete 8.6--10.2 pp more often than a single one, headroom for harnesses that exploit the non-determinism of LLM agents and can discriminate non-trivial from trivial differences between tests.

We make the following contributions:
\begin{itemize}[leftmargin=*]
    \item We formulate repository-level requirement-to-test generation as a requirement-driven, interface-assisted task in which an LLM agent must synthesize one complete, potentially multi-file Test Implementation while the target implementation is withheld.

    \item We construct and release \tool, comprising 5,086 authentic, execution-validated Requirement--Test Implementation instances from 8 projects in 5 programming languages, and measure behavioral adequacy relative to the developers' own tests.
    \item We evaluate eight agent--model--reasoning configurations that differ in the agent scaffold, the LLM, or the reasoning effort, uncover sharply diminishing returns and a persistent information gap, and identify promising directions for future research.
\end{itemize}

The code, data, and leaderboard are available in our repository~\cite{req2testbenchwebsite}.

\section{Related Work}

\definecolor{benchyes}{HTML}{1B7F3A}
\definecolor{benchno}{HTML}{B42318}
\providecommand{\yesmark}{\textcolor{benchyes}{\ding{51}}}
\providecommand{\nomark}{\textcolor{benchno}{\ding{55}}}
\providecommand{\partialmark}{\textcolor{benchno}{\bcancel{\ding{51}}}}

\begin{table}[t]
  \centering
  \caption{Comparison of resources pairing requirement-like inputs with ground-truth test code. 
  ``Source'' indicates whether an artifact was preserved from the subject project.
  ``Hierarchy'' indicates test levels and explicit requirement relations.}
  \label{tab:req2test-comparison}
  \scriptsize
  \setlength{\tabcolsep}{5pt}
  \renewcommand{\arraystretch}{1.16}

\begin{tabular}{llcccrc}
\toprule
\multicolumn{1}{l}{\multirow{2}{*}{\textbf{Resource}}}                 & \multirow{2}{*}{\textbf{Source}}                                                                      & \multirow{2}{*}{\textbf{Granularity}}                                                                                                                                        & \multirow{2}{*}{\textbf{\begin{tabular}[c]{@{}c@{}}Evaluation\\ Criteria\end{tabular}}} & \multirow{2}{*}{\textbf{Hierarchy}} & \multicolumn{2}{c}{\multirow{1}{*}{\textbf{Size}}}  \\ \cmidrule(lr){6-7}
\multicolumn{1}{c}{}                                               &          &                                                                    &                                                                                         &                                                                                           & \textbf{\#Pairs}                                       &  \textbf{\#PLs }                              \\ \midrule
E2EDev~\cite{liu2026e2edev}                  & Synthetic  & Scenario $\leftrightarrow$ Test Impl.                                                                                                             & N/A                                                                                     & N/A                                                                    & 703                                                        & 1                               \\ \midrule
GivenWhenThen~\cite{junior2026givenwhenthen} & Real & Feature $\leftrightarrow$ Test Files                                                                                                         & N/A                                                                                     & N/A                                                                    & 2,289                                                      & 3                               \\ \midrule
AutoGlue~\cite{shi2026autoglue}             & Real & Step $\leftrightarrow$ Method                                                                                                           & Ref.-based                                                                              & N/A                                                                    & 1,307                                                      & 1                               \\ \midrule
DevBench~\cite{li2024devbench}                                                                            & Manual  & Project $\leftrightarrow$ Test Files                                                      & Prog.-based                                                                           & N/A                                                                    & 22                                                         & 4                               \\ \midrule
\rowcolor{black!8}
\textbf{\tool}                                       & \textbf{Real} & \textbf{Scenario} $\leftrightarrow$  \textbf{Test Impl.}      & \textbf{Prog. \& Ref.-based}                                                               & \yesmark                                                                   & \textbf{5,086}                                             & \textbf{5}                      \\ \bottomrule
\end{tabular}
\end{table}

\textbf{Test-generation benchmarks.}
Existing test-generation benchmarks usually derive the test objective from prescribed code, a defect context, or an observed application.
Methods2Test~\cite{tufano2022methods2test} mines focal-method--JUnit-test pairs, while TestBench~\cite{zhang2024testbench} and TestGenEval~\cite{jain2025testgeneval} evaluate class- or file-level test generation using execution, coverage, and fault-detection or mutation metrics.
SWT-Bench~\cite{mundler2024swt} asks for a test that reproduces an issue report, and AutoE2E~\cite{alian2025feature} infers features by exploring a
running web application before generating end-to-end tests.
These works provide important execution-based protocols, but their task
boundary is supplied by focal code, a code-change or fault context, or an
observed application.
\tool studies the complementary requirement-defined setting, in which a
developer-authored Requirement determines the target behavior and an agent
must realize it as a complete, project-compatible Test Implementation.

\noindent\textbf{Requirement--test resources.}
\autoref{tab:req2test-comparison} compares the closest works that associate requirement artifacts with test code.
E2EDev~\cite{liu2026e2edev} provides LLM-generated requirements, BDD
Scenarios, and Step Definitions for web applications.
GivenWhenThen~\cite{junior2026givenwhenthen} mines developer-authored Feature
and Step Definition files at scale, but its scenario records contain
file-level artifacts rather than an execution-validated, Scenario-specific
Test Implementation.
DevBench~\cite{li2024devbench} includes benchmark-authored PRDs and executable
acceptance- and unit-testing tasks with mixed-provenance tests, but does not
map individual requirement criteria to individual ground-truth tests.
\tool instead recovers developer-authored Feature--Scenario artifacts and
their complete ground-truth Test Implementations from native BDD runtime
bindings, validates each mapping through positive and negative executions,
and uses the mapping for ground-truth-relative API, coverage, and mutation
comparisons.
Thus, \tool contributes the execution-validated, Scenario-level
Requirement--Test Implementation mapping and behavior-aware evaluation needed
to study complete Scenario realization, complementing prior work on
Gherkin-to-code generation.

\noindent\textbf{Requirements-based and BDD test generation.}
Requirements-based testing has long derived tests from structured or
natural-language specifications.
UMTG~\cite{wang2015umtg} derives executable acceptance tests from use cases
and a domain model, while Specmate~\cite{fischbach2020specmate} and
CiRA~\cite{fischbach2021cira} derive test designs from acceptance criteria or
conditional relations.
Recent LLM-based systems generate executable artifacts from requirements or
intermediate behavioral specifications.
AutoGlue~\cite{shi2026autoglue} generates one Java glue method for a target
Gherkin Step using its Feature and Scenario context together with retrieved
project artifacts.
Ferreira et al.~\cite{ferreira2025acceptance} generate Gherkin Scenarios and
Cypress tests from user stories and application HTML, whereas
AToMIC~\cite{fonseca2025atomic} generates Gherkin Scenarios, Page Objects,
and mobile UI tests from JIRA requirements and code changes.
Other systems use DSL requirements, validation intentions, or
product-requirement knowledge to guide testing around prescribed
architectures or focal
code~\cite{francca2026req2test,qi2026generatingprojectspecifictestcases,IntUT,yang2026uncoveringbusinesslogicbugs}.
These studies establish the value of requirements and project knowledge, but
differ in generation unit and task boundary.
Among these, AutoGlue is the closest task-level neighbor: it maps one Step to
one glue method, whereas \tool evaluates the complete, potentially multi-file
Test Implementation required to automate one Scenario.
Other systems target
specialized UI-automation pipelines or guide test generation around
prescribed architectures or focal implementations.

\section{\tool}
\label{sec:bench}

\tool recovers execution-validated pairs of Requirements and Test Implementations from developer-authored BDD artifacts.
This section defines the artifact meta-model, dataset construction, and evaluation criteria.

\begin{figure}
    \centering
    \includegraphics[width=1.0\linewidth]{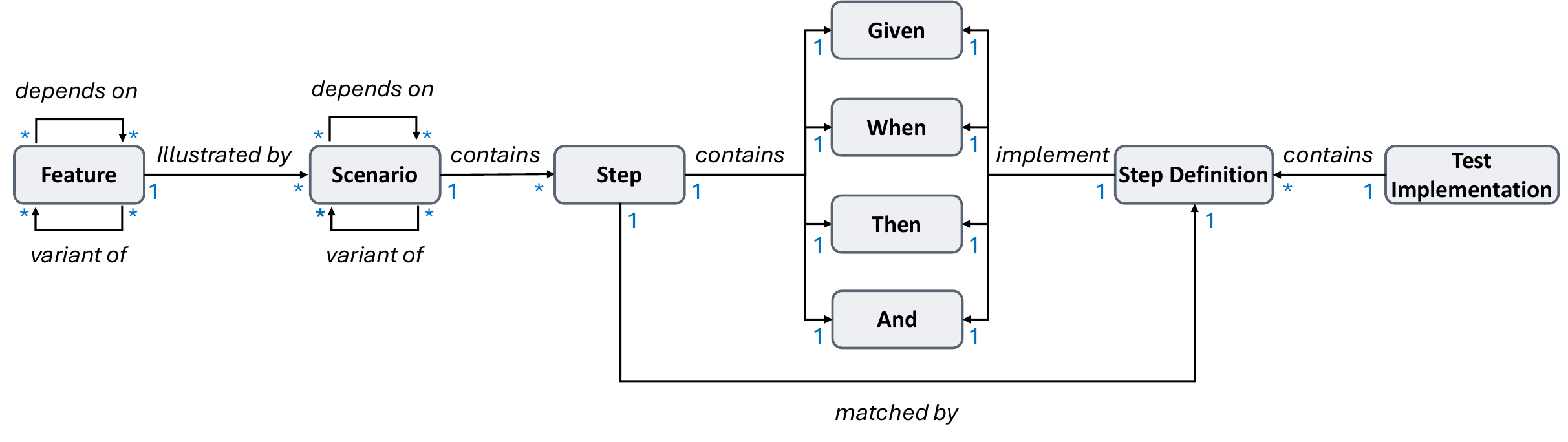}
    \caption{BDD-grounded meta-model of \tool. 
    A Requirement pairs an enclosing Feature with a target Scenario.
    A \texttt{Feature} is illustrated by \texttt{Scenario}s.
    Each \texttt{Scenario} contains an ordered sequence of Steps matched to executable Step Definitions and is automated by one complete, potentially multi-file Test Implementation. 
    A Test Implementation contains its matched Step Definitions and their helper code.
    Dependency and variant relations may connect Features or Scenarios.}
    \label{fig:meta-model}
    \vspace{-18pt}
\end{figure}

\subsection{Requirement-to-Test Artifact Meta-Model}
\label{sec:meta-model}

Each instance $i$ supplies the task inputs of \autoref{eq:req2test}, namely the Requirement $r_i=\langle F_i,S_i\rangle$, the target interfaces $\mathcal{I}_i^{\mathrm{tar}}$, and both views of the project snapshot $P_i$.
It also holds evaluator-only data: the developer-authored ground-truth Test Implementation $T_i^{\mathrm{gt}}$ and its API calls, coverage, and mutant kills, against which a generated Test Implementation $T_i^{\mathrm{gen}}$ is compared.

\noindent\textbf{Entity.}
\autoref{fig:meta-model} shows the BDD-grounded meta-model:
\stepcircle{1} A \emph{Feature} provides a developer-authored capability description and groups related \emph{Scenarios}.
\stepcircle{2} Each Scenario expresses a concrete behavioral example through ordered \emph{Steps}.
\stepcircle{3} Steps
\textit{Given}, \textit{When}, and \textit{Then} conventionally express context, an action or event, and an expected outcome.
An \textit{And} continues the preceding role~\cite{cucumberGherkin}.
Each effective Step in a retained instance binds uniquely to an executable \emph{Step Definition}.
\stepcircle{4} A \emph{Test Implementation} is the project-local test-side closure needed to bind, execute, and check the target Scenario: its matched Step Definitions and the helper code that they require.
Each Test Implementation has a \emph{test level}, namely component, integration, or system testing.

\noindent\textbf{Relation.}
The meta-model defines five relation types.
\stepcircle{1}
\texttt{illustrated\_by} links a Feature to its Scenarios;
\stepcircle{2}
\texttt{contains} links a Scenario to its ordered Steps and a Test Implementation to its constituent test-side artifacts;
\stepcircle{3} \texttt{matched\_by} links a Step to its selected Step Definition.
Multiple Steps may use the same Step Definition, and Test Implementations may overlap.
\stepcircle{4} The optional, directed \texttt{depends\_on} relation records prerequisite or decomposition relations between Features or between Scenarios.
\stepcircle{5} The symmetric \texttt{variant\_of} relation connects artifacts expressing the same core intent under different inputs, preconditions, configurations, or expected outcomes.

\subsection{Dataset Construction}
\label{sec:dataset-construction}
We construct \tool in five stages.
Stage~I assembles candidate repositories and their BDD artifacts.
Stage~II selects projects for execution and filters their Scenarios.
Stage~III recovers and validates Requirement--Test Implementation mappings, identifies target interfaces, and constructs the agent views.
Stage~IV constructs the mutation evidence and establishes mutation scorability. Stage~V labels test levels and requirement relations.

\noindent\textbf{Stage I: Candidate discovery and artifact collection.}
We assembled candidate GitHub repositories containing developer-authored Gherkin \texttt{.feature} files and BDD test code, using framework dependencies, configuration files, and Step Definitions as static evidence.
\textbf{Four execution frameworks} are Cucumber (Java and Ruby), Behave (Python), Behat (PHP), and Godog (Go).
An initial survey of the 100 highest-starred discovered repositories with more than 100 recorded stars identified potentially usable Scenarios with static Step Definition matches in 46 repositories.
Further discovery added candidates beyond this survey.
For candidates taken forward, we pinned revisions and collected feature files, Step Definitions, framework configuration, hooks, fixtures, test data, and helpers.
This stage yields candidate repositories and their artifacts for execution checks in Stage~II.

\noindent\textbf{Stage II: Project admission and Requirement filtering.}
We favored higher recorded GitHub star counts, more recent last-push dates, and larger BDD suites, measured by their numbers of feature files and Scenarios.
These preferences were balanced against domain and testing diversity.
The eight selected projects span \textbf{five programming languages} and cover \textbf{six domains}: e-commerce (CoreShop, PrestaShop, and Sylius), document processing (python-docx), a blockchain SDK (Algorand SDK), IoT services (Kapua), an online petitions service (e-petitions), and agent software (coddy-agent).
We sought both direct library tests and multistep workflows involving shared fixtures, database state, and service/API interactions.

We removed empty, comment-only, and unparsable feature files and Scenarios that need a browser or a live service, and retained Features with nonempty titles and Scenarios with nonempty Step sequences.
Applicable \texttt{Background} Steps are included in execution order, forming the Scenario's \emph{effective Steps}.
A Scenario Outline remains one instance: its required \texttt{Examples} rows are executed after parameter substitution and combined into one instance-level outcome.
We provisioned each project at its pinned revision and invoked its native BDD runner.
A Scenario was retained only if the project initialized, every effective Step bound uniquely to a non-pending Step Definition, and the Scenario passed in isolation across three runs.
We excluded build or initialization failures, undefined or ambiguous Steps, timeouts, and unstable outcomes without rewriting project-authored requirements or tests to make them pass.
This execution gate follows repository-scale benchmark construction practice~\cite{jimenez2024swebench,liu2026e2edev}.

\noindent\textbf{Stage III: Test-Implementation recovery, target-interface identification, and agent views.}

\textbf{Test Implementation Recovery.}
For each retained Scenario, we recorded the Step Definition selected by the runner for every effective Step.
Starting from these runtime bindings, we recovered the transitive project-local test-side closure, including helpers, but excluding production code, third-party framework code, and the testing infrastructure that the whole suite shares.
Shared definitions and supporting artifacts can belong to more than one closure.
We validated each candidate closure in two derived workspaces sharing the same production revision, execution command, configuration, and project-global BDD infrastructure.
Among project-local test-side artifacts, the positive workspace retains only the recovered closure and invariant infrastructure, and the target Scenario must pass.
In the negative workspace, which withholds the closure, the runner must still initialize, but the target Scenario must not pass.
Both outcomes must hold across three runs.
Only validated closures become ground-truth Test Implementations $T_i^{\mathrm{gt}}$.

\textbf{Target Interface Identification.}
We identify target interfaces by \emph{Step-attributed direct-call tracing} of ground-truth tests.
We select Steps \textit{When} and \textit{Then}, excluding Background and other context Steps.
We retain public production methods called directly from test-side code during these Steps, excluding transitive production calls and constructors invoked only from check Steps.
Method declarations are deduplicated and grouped by declaring type to obtain $\mathcal{I}_i^{\mathrm{tar}}$.

This restriction reduces setup-related disclosure while retaining concrete interaction and observation points.
Only declarations are supplied, not ground-truth call arguments, invocation order, assertions, or test code.
The result is an observed testing boundary, not an exhaustive semantic localization.
Instances with an empty set are retained, in total 234 instances.
Their action and check Steps call no production method directly, e.g., an e-petitions test acts through an in-process HTTP client, and excluding them would remove 85 of the 98 e-petitions instances.

\textbf{Agent Views.}
For target-hidden, a source-level mapping uses the Requirement, ground-truth bindings, and pinned code and configuration to identify the bodies realizing the target behavior, including relevant delegation but not unrelated supporting infrastructure.
The interface list alone is not the body-removal list.
Language-aware transformations replace mapped bodies while preserving declarations, and all-hidden applies them throughout the declared production scope.

We then materialize $C_{\mathrm{TH}}(P_i)$ and $C_{\mathrm{AH}}(P_i)$ with the same Requirement, interface annotations, permitted contract information, and retained infrastructure.
Both views remove the Step Definitions and helpers of all Scenarios. Both keep the testing infrastructure that the whole suite shares, such as hooks, fixtures, and test data.
Neither shows runtime traces, coverage, or mutation records.

\noindent\textbf{Stage IV: Mutation data and scorability.}
We implement established mutation-operator families as language-specific source transformations~\cite{jia2011mutation,pit_mutators,infection_mutators}.
We keep mutant generation separate from test execution, so that the ground-truth and generated tests can run against the same recorded mutants using each project's native BDD runner.
Operator families include conditional-boundary changes, negation, arithmetic and logical-operator replacement, constant and return-value replacement, and method-call removal.
The supported operators and language-specific adaptations are recorded per project rather than assumed identical across languages.
Mutants are restricted to the declared production scope, excluding test scaffolding, generated, third-party, and other out-of-scope code.
The versioned manifest records each mutant's source location, change, and stable identity.

Let $M_{P_i}$ denote the selected production-mutant manifest and $C_i^{\mathrm{gt}}$ the production file--line locations covered by $T_i^{\mathrm{gt}}$ on the unmodified snapshot.
For the source locations $\mathcal{L}(m)$ assigned to mutant $m$ by the project's fixed coverage mapping, the instance pool is
\begin{equation}
U_i^{\mathrm{mut}}=
\left\{m\in M_{P_i}\mid
\mathcal{L}(m)\cap C_i^{\mathrm{gt}}\neq\emptyset\right\}.
\label{eq:mutant-pool}
\end{equation}
Thus, pool membership depends on the ground-truth test's coverage.

We execute $T_i^{\mathrm{gt}}$ separately against every mutant in $U_i^{\mathrm{mut}}$ and record an outcome for each instance--mutant pair.
Pair-level evidence is necessary because one project mutant can occur in multiple instance pools and have different outcomes for different ground-truth tests.
For a test passing on the unmodified snapshot, a mutant is killed when its execution produces a test failure under the declared outcome policy, and it survives when the execution passes.
A ground-truth run that times out, or fails before any Step executes, is an infrastructure outcome, and its pair is excluded.
The valid ground-truth kills form $K_i^{\mathrm{gt}}$.
Furthermore, mutants killed by the ground-truth test but survived by its assertion-stripped counterpart form an \emph{assertion-dependent subset} for secondary analysis of the removed checks.

\noindent\textbf{Stage V: Test levels and requirement relations.}
Stage~V labels test levels and requirement relations, and neither label changes the task.

\textbf{Test levels.}
We label a ground-truth Test Implementation by how its Step Definitions reach the production code: 
\stepcircle{1}~\emph{component}-level when they instantiate production classes directly; 
\stepcircle{2}~\emph{integration}-level when they obtain real services from the application's own wiring, such as a dependency-injection container or a command bus; and 
\stepcircle{3}~\emph{system}-level when they drive the assembled application only through its external interface, such as its command-line entry point or the documented entry point of a library.
Among our collected test implementations, 4,276 tests are named by the official developers, e.g., Kapua's separate unit and integration runners, and our labels agree for 4,266 of these 4,276 instances.
None is end-to-end test, as Stage~II excludes Scenarios that need a browser.

\begin{table*}[t]
\centering
\caption{
Statistics of \tool.
Left: totals over all instances.
Right: minimum, maximum, and mean per instance.
C / I / S: component, integration, and system level.
A ground-truth kill is an instance--mutant pair whose mutant the ground-truth test kills.
}
\label{tab:dataset-statistics}
\begingroup
\footnotesize
\setlength{\tabcolsep}{3.5pt}
\renewcommand{\arraystretch}{1.05}

\begin{minipage}[t]{0.4\textwidth}
\vspace{0pt}
\centering
\begin{tabularx}{\linewidth}
{@{}>{\raggedright\arraybackslash}Xr@{}}
\toprule
\textbf{Statistic} & \textbf{Total} \\
\midrule
\rowcolor{black!8}
\multicolumn{2}{@{}l@{}}{\textit{Benchmark}} \\
Projects & 8 \\
Programming languages & 5 \\
BDD frameworks & 4 \\
\midrule
Instances & 5,086 \\
Features & 1,415 \\
Multi-file Test Impl. & 1,741 \\
\midrule
Test levels (C / I / S) & 588 / 2,066 / 2,432 \\
\texttt{depends\_on} pairs & 634 \\
\texttt{variant\_of} pairs & 3,584 \\
\midrule
Mutation operators & 11 \\
Ground-truth kills & 247,585 \\
\bottomrule
\end{tabularx}
\end{minipage}
\hfill%
\begin{minipage}[t]{0.55\textwidth}
\vspace{0pt}
\centering
\begin{tabularx}{\linewidth}
{@{}>{\raggedright\arraybackslash}Xrrr@{}}
\toprule
\textbf{Statistic} &  \textbf{Min} & \textbf{Max} & \textbf{Avg.} \\
\midrule
\rowcolor{black!8}
\multicolumn{4}{@{}l@{}}{\textit{Feature}} \\
Scenarios per Feature & 1 & 64 & 4.75 \\
Words per Feature & 1 & 151 & 28.97 \\
\rowcolor{black!8}
\multicolumn{4}{@{}l@{}}{\textit{Scenario}} \\
Steps per Scenario & 2 & 77 & 10.96 \\
Words per Scenario & 11 & 762 & 110.30 \\
\rowcolor{black!8}
\multicolumn{4}{@{}l@{}}{\textit{Test Impl.}} \\
Files per Test Impl. & 1 & 25 & 5.29 \\
Step defs.\ per Test Impl. & 1 & 36 & 7.37 \\
Lines per Test Impl. & 37 & 8,681 & 1,763.50 \\
\rowcolor{black!8}
\multicolumn{4}{@{}l@{}}{\textit{Mutation}} \\
Ground-truth kills per instance & 2 & 588 & 120.71 \\
Assertion-dependent kills per instance & 0 & 258 & 34.05 \\
\bottomrule
\end{tabularx}
\end{minipage}

\endgroup
\vspace{-10pt}

\end{table*}

\textbf{Requirement relations.}
Scenario $A$ \texttt{depends\_on} Scenario $B$ when $A$ performs the behavior that $B$ checks and relies on it, as a precondition or as one of its parts.
For example, a Kapua Scenario that tests changing a user's password first creates a password credential in a \textit{Given} Step, and another Scenario performs the same creation and checks that it succeeds.
As developers write such shared behavior either in a \texttt{Background} or among a Scenario's own Steps, we look in both: $A$'s \textit{Given} and \textit{When} Steps must run every Step Definition that $B$ runs before its checks.
An LLM judge (DeepSeek-V4.1-Flash) then removes pairs in which $A$ reuses only setup that $B$ does not itself check.
Scenarios whose \textit{Given} and \textit{When} Steps run the same Step Definitions, differing only in data or expected outcomes, are \texttt{variant\_of} each other.
This yields 634 \texttt{depends\_on} relations and 3,584 \texttt{variant\_of} pairs.

\noindent\textbf{Benchmark statistics.}
Stages~I--V yield 5,086 execution-validated instances from 1,415 Features in eight projects.
\autoref{tab:dataset-statistics} summarizes them, and \autoref{tab:dataset-statistics-break} breaks them down by project.

\begin{table}[t]
\centering
\caption{Project-level breakdown of \tool: instances, per-instance measurements of the Requirement and its ground-truth Test Implementation, and mutation data. Means are over the evaluated instances. 
}
\label{tab:dataset-statistics-break}
\footnotesize
\setlength{\tabcolsep}{2pt}
\renewcommand{\arraystretch}{1.05}
\begin{tabular*}{\linewidth}{@{\extracolsep{\fill}}ll*{11}{r}@{}}
\toprule
 & & \multicolumn{3}{c}{\textbf{Instances}}
   & \multicolumn{5}{c}{\textbf{Mean per evaluated instance}}
   & \multicolumn{3}{c}{\textbf{Mutation}} \\
\cmidrule(lr){3-5}\cmidrule(lr){6-10}\cmidrule(lr){11-13}
\textbf{Project} & \shortstack[l]{\textbf{Language} /\\\textbf{framework}}
 & \textbf{Valid}. & \textbf{Eval.} & \shortstack[r]{\textbf{Step}\\\textbf{defs.}}
 & \textbf{Steps} & \shortstack[r]{\textbf{Target}\\\textbf{I/F}} & \textbf{Files} & \textbf{Lines} & \textbf{API}
 & \textbf{Mutants} & \textbf{Pairs} & \shortstack[r]{\textbf{Kill}\\(\textbf{\%})} \\
\midrule
coddy-agent & Go, godog & 282 & 274 & 835 & 5.8 & 3.3 & 2.5 & 1,357 & 8.6 & 20,243 & 227,886 & 24.8 \\
Algorand SDK & Java, Cucumber & 117 & 111 & 188 & 4.1 & 6.2 & 2.2 & 542 & 7.5 & 7,766 & 12,853 & 58.4 \\
Kapua & Java, Cucumber & 605 & 330 & 415 & 6.3 & 9.5 & 1.5 & 2,286 & 18.9 & 48,911 & 33,000 & 41.8 \\
CoreShop & PHP, Behat & 422 & 334 & 371 & 16.1 & 3.9 & 13.8 & 2,631 & 44.0 & 10,205 & 171,174 & 28.2 \\
PrestaShop & PHP, Behat & 1,352 & 348 & 713 & 21.5 & 13.8 & 8.2 & 2,911 & 35.0 & 71,670 & 34,355 & 29.4 \\
Sylius & PHP, Behat & 1,950 & 319 & 552 & 11.9 & 1.9 & 3.4 & 1,215 & 2.0 & 12,373 & 256,324 & 26.1 \\
python-docx & Python, behave & 239 & 237 & 466 & 2.9 & 3.1 & 2.1 & 368 & 6.0 & 2,225 & 40,359 & 74.4 \\
e-petitions & Ruby, Cucumber & 119 & 98 & 166 & 10.6 & 0.2 & 4.0 & 652 & 0.4 & 5,997 & 105,193 & 13.7 \\
\midrule
\textit{Total} & & 5,086 & 2,051 & 3,706 & 11.0 & 6.0 & 5.3 & 1,763 & 18.7 & 179,390 & 881,144 & 28.1 \\
\bottomrule
\end{tabular*}
\vspace{-12pt}
\end{table}

\subsection{Evaluation Criteria}
\label{sec:eval-metrics}
A generated test must first be a valid, executable implementation and must then exercise the behavior required by the scenario.  
We therefore report two complementary layers of metrics:
\textit{stage-based execution outcomes} that identify where generation or evaluation fails, and \textit{ground-truth-relative behavioral comparison} that provides evidence beyond a passing execution.

\tool saves the generated test files in a separate copy of the project containing the complete production implementation.
\tool then executes the target Scenario using the project's native BDD runner and records its outcome and behavioral measurements.
For each agent and view, \tool evaluates one final generated test per instance in a predeclared population of size $N$.
Intermediate edits, local checks, and repairs belong to the same generation run and do not count as additional evaluation candidates.

\subsubsection{Stage-based execution outcomes}
For each instance with a recorded result, we report success or the first stage at which test generation or evaluation fails.
Failures caused by benchmark infrastructure are recorded separately and excluded from the following categories:
\begin{itemize}[leftmargin=*]
    \item \textbf{Generation failure:} the agent returns no usable test code, no writable file, or an invalid submission format.
    \item \textbf{Build failure:} the generated test files cannot be parsed,
    compiled, imported, or initialized with the project's test environment.
    This category also covers interpreted languages.  
    \item \textbf{Binding failure:} the test framework cannot bind every Scenario step to a Step Definition, or encounters an undefined, ambiguous, or pending step.
    \item \textbf{Execution failure:} the scenario does not complete due to an exception, crash, or timeout, without an assertion-failure outcome.
    \item \textbf{Assertion failure:} the generated test runs and an assertion fails against the unmodified production implementation.
    \item \textbf{Pass:} the selected Scenario completes with every effective step passing and the test command exits successfully.
\end{itemize}

Let $\mathcal{E}$ be the attempted instances with recorded outcomes in the six categories above, after documented infrastructure exclusions, $N_{\mathrm{exec}}=|\mathcal{E}|$, and $o_i$ the outcome of instance $i$.
For $N_{\mathrm{exec}}>0$, the execution success rate is
\begin{equation}
  \mathrm{Pass@1}=\frac{1}{N_{\mathrm{exec}}}\sum_{i\in\mathcal{E}}
  \mathbf{1}[o_i=\mathrm{pass}].
\end{equation}
All failure rates use the same denominator $N_{\mathrm{exec}}$.
We report $N$, $N_{\mathrm{exec}}$, and the number and reasons for excluded instances.

\subsubsection{Ground-truth-based behavioral comparison.}
$\mathrm{Pass@1}$ captures whether a generated test executes successfully, but it does not quantify the extent to which the test exercises and checks the behavior required by the target Scenario. In the extreme, a Scenario may pass even when all generated Step-Definition bodies are empty. 

Therefore, \tool further compares the passing generated test with its developer-written ground-truth test along three dimensions: 
\ding{172}~API-call F1 and \ding{173}~coverage Jaccard describe the exercised behavior, while \ding{174}~mutant-kill recall assesses fault detection.
Our primary end-to-end measure combines execution success with complete mutant-kill recall.
For each dimension $s$, let $GT_i^s$ and $Gen_i^s$ denote the ground-truth and generated-test sets.

\noindent\textbf{API-call F1.}
The API sets identify public production methods directly called by the ground-truth or generated Test Implementation, including setup, action, and check Steps.
Production-to-production calls and calls made only by retained testing infrastructure are excluded.
For $GT_i^{\mathrm{api}}\neq\emptyset$,
\begin{equation}
  \mathrm{APIF1}_i=2\times
  \frac{|Gen_i^{\mathrm{api}}\cap GT_i^{\mathrm{api}}|}
       {|Gen_i^{\mathrm{api}}|+|GT_i^{\mathrm{api}}|}.
\end{equation}

\noindent\textbf{Coverage Jaccard.}
Coverage compares sets of executed code lines across all production files:
\begin{equation}
  \mathrm{CovJaccard}_i=
  \frac{|Gen_i^{\mathrm{cov}}\cap GT_i^{\mathrm{cov}}|}
       {|Gen_i^{\mathrm{cov}}\cup GT_i^{\mathrm{cov}}|}.
\end{equation}
High coverage overlap does not establish that the required outcomes are asserted.

\noindent\textbf{Mutant-kill recall.}
Within the fixed pool $U_i^{\mathrm{mut}}$, $GT_i^{\mathrm{mut}}$ is $K_i^{\mathrm{gt}}$ restricted to retained instance--mutant pairs, and $Gen_i^{\mathrm{mut}}$ contains the generated test's kills on those same pairs.
Timeout and infrastructure exclusions are reported with counts.
All ground-truth kills in the scored scope are used, not only the assertion-dependent subset.
For $GT_i^{\mathrm{mut}}\neq\emptyset$,
\begin{equation}
  \mathrm{MutRecall}_i=
  \frac{|Gen_i^{\mathrm{mut}}\cap GT_i^{\mathrm{mut}}|}
       {|GT_i^{\mathrm{mut}}|}.
\end{equation}

\noindent\textbf{End-to-end success.}
Let $\mathcal{I}_{\mathrm{mut}}$ be the planned instances with supported, nonempty ground-truth kill sets.
We define ground-truth-kill-complete pass@1 as
\begin{equation}
  \mathrm{KillComplete@1} =
  \frac{1}{N}
  \sum_{i\in\mathcal{I}_{\mathrm{mut}}}
  \mathbf{1}\left[o_i=\mathrm{pass}\ \wedge\ \mathrm{MutRecall}_i=1\right].
\end{equation}
The denominator is the complete planned population $N$, and a non-passing instance is never kill-complete.
A requirement's text may leave project-specific setup and expected outcomes implicit.
We therefore use the developer-written ground-truth test as a concrete reference for evaluating behavioral adequacy (Definition~2).
A passing test that misses any evaluated ground-truth kill is \emph{kill-incomplete}.
It may still satisfy the requirement's text under a defensible interpretation, but it misses faults that the developers' test detects, so it is not adequate in the project's context.

\noindent\textbf{Aggregation and interpretation.}
Within each project, we average API F1, coverage Jaccard, and mutant-kill recall over all $N$ instances (API F1 over those whose ground truth calls a production method directly), with a non-passing test scoring 0, so that these means are end-to-end measures like Pass@1 and KillComplete@1.
Project scores are then averaged with equal weight.
These means describe each agent's passing tests, whereas $\mathrm{KillComplete@1}$ uses the planned population.
None of these comparisons establishes semantic equivalence or exhaustive requirement satisfaction.

\subsubsection{Efficiency.}
For each configuration and view, we report generation time, LLM calls, input and output tokens, and monetary cost where recorded.
Costs of post-generation test execution and offline mutation analysis are reported separately.
In addition to per-instance summaries and totals, we report generation cost per execution pass and per ground-truth-kill-complete pass.

\section{Experimental Design}
\label{sec:exp-design}

In this section, we
formulate the research questions, introduce the evaluated approaches,
specify how each approach is executed, and describe how the resulting
observations are analyzed.

\subsection{Research Questions}

Our empirical study addresses the following research questions: RQ1 and RQ2 answer question Q1 of the introduction, RQ3 and RQ4 question Q2, and Section \ref{sec:probe} question Q3.

\noindent\textbf{RQ1 (Effectiveness):}
\emph{How effectively do LLM agents solve Req2Test tasks in \tool?}
This question investigates the current capability of state-of-the-art LLM agents to transform natural-language requirements into executable and behaviorally adequate tests.

\noindent\textbf{RQ2 (Efficiency).}
\emph{How efficiently do the LLM agents solve the Req2Test
tasks, and what effectiveness--efficiency trade-offs do they exhibit?}
We measure monetary cost, end-to-end time, model requests, and tokens per generation, and relate cost to effectiveness through the cost per kill-complete pass.

\noindent\textbf{RQ3 (Associated Factors).}
\emph{What requirement-, test-, and repository-level characteristics are associated with successful Req2Test generation?}
This question investigates whether current agents perform consistently across the heterogeneous settings represented by \tool.
We compare the all-hidden and target-hidden views on the same instances, and relate success to characteristics of the Requirement, its ground-truth Test Implementation (including its test level), the hidden code, and the Scenario's prerequisites among the \texttt{depends\_on} relations. 

\noindent\textbf{RQ4 (Failure Modes).}
\emph{What failure modes do current agents exhibit, and what limitations do these failures reveal?}
We first identify the evaluation stage at which an unsuccessful generation fails.
We then derive failure modes bottom-up from a stratified sample, explaining both why failing tests fail and why passing tests miss ground-truth kills.

\subsection{Evaluated Models and Agents}
We evaluate eight agent--model--reasoning configurations using four LLM backbones: \ding{172}~DeepSeek-V4.1-Flash (DS)~\cite{deepseekai2026deepseekv41flashpushinglimitskv}, \ding{173}~GPT-6 Luna~\cite{gpt6}, \ding{174}~GPT-6 Sol~\cite{gpt6}, and \ding{175}~Claude Sonnet 5.5~\cite{claudesonnet55}.
The four general-purpose coding agents are \ding{172}~\textbf{OpenHands} (openhands-sdk \texttt{1.34.0})~\cite{wang2025openhandsopenplatformai}, \ding{173}~\textbf{Aider} (\texttt{0.86.2})~\cite{aider}, \ding{174}~\textbf{Codex CLI} (\texttt{0.153.4}; \texttt{0.157.1} for GPT-6 Sol)~\cite{codex}, and \ding{175}~\textbf{Claude Code} (\texttt{2.1.274})~\cite{claude-code}.
Reasoning effort is set to \textit{disabled} or \textit{high}, depending on the configuration.
We call each combination of agent, LLM, and reasoning effort a configuration and name it by its agent and LLM, e.g., Codex/DS.
Codex/DS~(off) runs DS without reasoning.
Claude Code/Sonnet runs Claude Sonnet 5.5 at high effort with an output limit of 128,000 tokens per request, and Claude Code/DS uses a limit of 32,000 tokens.

Aider and OpenHands run DS in non-thinking mode, with limits of 10 reflections and 200 LLM responses, respectively.
Codex CLI and Claude Code each allow up to 200 LLM requests.
Each agent receives the target requirement and corresponding
target interfaces, and uses its native mechanisms for repository
inspection, file editing, and command execution.
We retain the native interaction workflow of each agent rather than
forcing all agents into an identical implementation interface.

\subsection{Implementation Details}
\label{sec:impl-detail}
\noindent\textbf{Generation.}
Each generation receives the Requirement, the target interfaces, and one repository view inside a \texttt{bubblewrap} sandbox~\cite{bubblewrap} whose network reaches only the model provider, under a common 3,600\,s wall-clock limit.
Agents may inspect the view and run local commands, including their own tests wherever the sandbox can load the retained code, but no such run can exercise the target behavior: its production bodies are placeholders that fail with an explicit ``implementation hidden'' error, and the harness provisions no application runtime and returns no execution feedback.
The instructions state that the final test runs against a separate, complete implementation.
We collect the test files from the workspace diff and discard production edits.
Each instance receives one scored generation per configuration and view.
Generations interrupted by provider-side limits are re-run, and none is repeated because of its outcome.
To ensure that no generation observes ground-truth-derived artifacts, we audited the trajectories for access to such files and regenerated the affected generations in a hardened sandbox under a rule fixed in advance.

\noindent\textbf{Evaluation.}
The evaluator runs each final test once against the complete production code with the project's native runner in a network-isolated container (600\,s, coverage enabled), and runs every passing test against the instance's mutant pool (Section~\ref{sec:dataset-construction}).

Many Scenarios of a project are tested through the same Step Definitions, so their Test Implementations are near-duplicates.
We therefore evaluate one per group of instances whose ground-truth Test Implementations use the same Step-Definition methods.
All configurations run on these 2,051 instances from 780 Features in both views.

\noindent\textbf{Statistical analysis.}
Brackets indicate 95\% confidence intervals obtained by cluster
bootstrap resampling of feature files within each project.
For the paired comparisons, we compute
one-sided $p$-values using exact McNemar tests or paired sign-flip tests.

\section{Results}\label{sec:results}

\begin{table*}[t]
\centering
\caption{Overall effectiveness: means over eight projects. A non-passing test counts as 0 in similarity columns.}
\label{tab:overall-performance}
\begingroup
\footnotesize
\setlength{\tabcolsep}{2.5pt}
\renewcommand{\arraystretch}{1.05}

\providecommand{\resultpair}[5]{\multirow{2}{*}{#1} & \multirow{2}{*}{#2} & \multirow{2}{*}{#3}
  & Target-hidden & #4 \\
  & & & All-hidden & #5 \\
}

\begin{tabular*}{\textwidth}{@{\extracolsep{\fill}}llllccccc@{}}
\toprule
& & & & \multicolumn{2}{c}{\textbf{End-to-end success (\%)}}
& \multicolumn{3}{c}{\textbf{Similarity to ground truth (\%)}} \\
\cmidrule(lr){5-6}\cmidrule(lr){7-9}
\textbf{System} & \textbf{Model} & \textbf{Reasoning} & \textbf{View}
& \textbf{Pass@1}
& \textbf{KillComplete@1}
& \textbf{API $F_1$}
& \textbf{Coverage $J$}
& \textbf{Mutation $R$} \\
\midrule
\resultpair{Aider}{DS}{off}
  {9.7 & 4.5 & 6.6 & 8.0 & 8.5}
  {9.7 & 4.4 & 6.7 & 8.0 & 8.7}  \hline
\resultpair{OpenHands}{DS}{off}
  {32.7 & 18.1 & 26.4 & 30.9 & 31.2}
  {26.5 & 16.6 & 21.2 & 25.2 & 25.3} \hline
\resultpair{Claude Code}{DS}{high}
  {80.5 & 34.4 & 62.0 & 73.2 & 73.3}
  {73.7 & 30.7 & 57.4 & 67.9 & 67.1} \hline
\resultpair{Claude Code}{Sonnet 5.5}{high}
  {80.9 & 35.9 & 61.7 & 76.0 & 75.6}
  {78.4 & 34.8 & 59.1 & 74.0 & 73.7} \hline
\resultpair{Codex CLI}{DS}{off}
  {57.6 & 23.6 & 41.9 & 53.1 & 53.4}
  {47.4 & 21.3 & 34.2 & 44.0 & 44.0} \hline
\resultpair{Codex CLI}{DS}{high}
  {82.8 & 33.8 & 64.2 & 75.9 & 76.3}
  {75.4 & 33.7 & 56.5 & 70.0 & 69.1} \hline
\resultpair{Codex CLI}{GPT-6 Luna}{high}
  {66.2 & 28.3 & 49.9 & 59.6 & 60.6}
  {57.1 & 27.8 & 42.9 & 52.7 & 53.6} \hline
\resultpair{Codex CLI}{GPT-6 Sol}{high}
  {73.5 & 31.3 & 54.2 & 65.3 & 66.1}
  {65.3 & 29.1 & 48.2 & 59.7 & 60.3}
\bottomrule
\end{tabular*}
\endgroup
\vspace{-13pt}
\end{table*}

\subsection{RQ1: Effectiveness}\label{sec:rq1}

\begin{figure}[t]
  \centering
  \includegraphics[width=\linewidth]{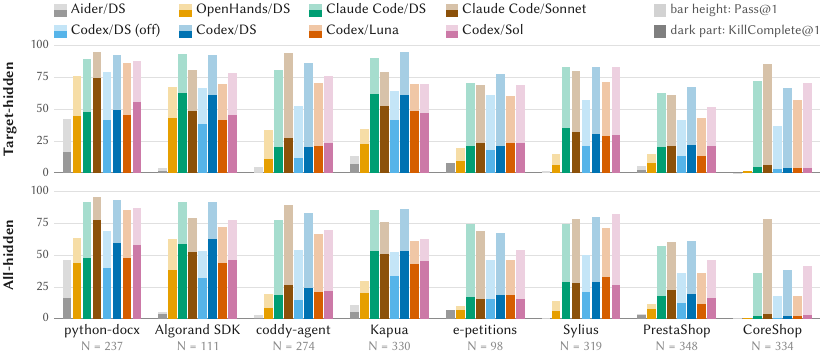}
  \caption{Pass@1 (bar height) and KillComplete@1 (dark part), in percent, per project, view, and configuration.}
  \label{fig:breakdown}
  \vspace{-10pt}
\end{figure}

\begin{figure}[t]
  \centering
  \includegraphics[width=\linewidth]{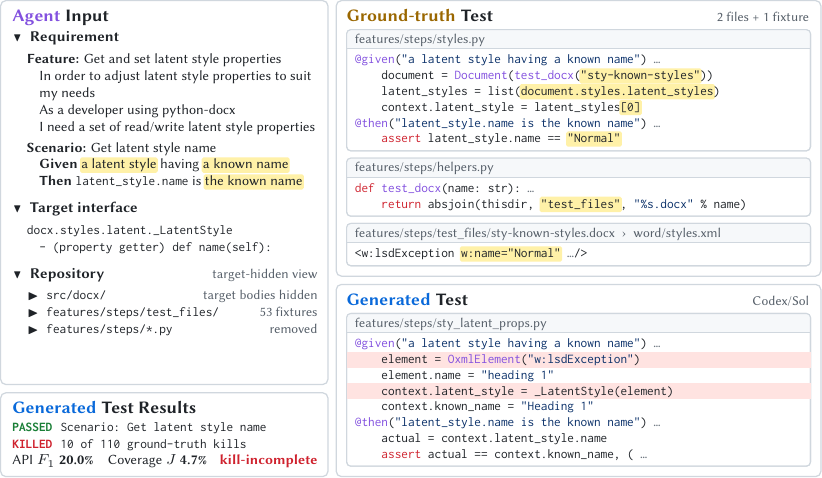}
  \caption{An illustrative example from \texttt{python-docx}, \texttt{sty-latent-props.feature:36} (target-hidden view). Left: the agent's input and its test result. 
  Right: the ground-truth test and the test generated by Codex/Sol. Yellow: what the Requirement leaves abstract and how the ground truth fills it; red: where the generated test constructs the object itself.}
  \label{fig:example}
  \vspace{-12pt}
\end{figure}

\noindent\textbf{Passing is common, but kill-completeness is not.}
In \autoref{tab:overall-performance}, the best-passing configuration, Codex/DS, passes 82.8\% of the target-hidden instances ([80.5, 84.8]) but is kill-complete on only 33.8\% ([28.4, 38.1]).
As shown in \autoref{fig:breakdown}, per project, only 6.7--66.2\% of the passing tests of all configurations and views are kill-complete.
The gap persists when the generation is repeated, although single verdicts vary (Section~\ref{sec:probe}).
A passing test of Codex/DS reproduces 91\% of its ground-truth test's kills on average, yet the gap between Pass@1 and KillComplete@1 is not merely a matter of demanding every kill.
Under a relaxed criterion that accepts a test reproducing at least 95\% of them, Codex/DS still succeeds on only 53.9\% of the target-hidden instances, 28.9 percentage points (pp) below its Pass@1.
High coverage similarity does not imply kill-completeness either.
On the six projects other than coddy-agent and Kapua, 17.4--69.7\% of the passing tests with a coverage Jaccard of at least 0.9 are kill-incomplete.
\autoref{fig:example} shows a concrete example: the test of Codex/Sol passes, but it constructs the latent style directly in its own code instead of loading it from the project's fixture file, and kills only 10 of the 110 mutants that the ground-truth test kills.

\noindent\textbf{The gap persists when only check-dependent kills count.}
We empty the bodies of the ground-truth tests' \textit{Then} Step Definitions to identify mutants killed by the original tests but surviving the ablated tests.
Across the seven DS and GPT-6 configurations, counting only these kills reduces the gap between Pass@1 and KillComplete@1 by only 24--37\%, without changing the configurations' KillComplete@1 ranking.
When counting only kills inside the hidden target implementation, the gap also persists, e.g., 11.9~pp for  Codex/DS.

\noindent\textbf{Kill-incomplete tests mostly fall short on checking, not only on the route.}
We find agents fall short on both clauses of Definition~2, i.e., ``exercises the behavior'' and ``checks expected outcomes''.
For example, only 43.5\% of the target-hidden instances get a passing Codex/DS test that executes every mutant that the developers' test kills, and only 41.1\% get one that kills every such mutant it executes.
Of its kill-incomplete tests, 77.9\% execute such a mutant and still pass, and only 22.1\% miss only mutants on lines they never run.
Thus, most kill-incomplete tests run the faulty code on their own path and check less than the developers' test there.

\noindent\textbf{Scaffold and reasoning effort matter, but the costliest LLM is not the best.}
\stepcircle{1}~\textit{Agent scaffold}: with DS and reasoning off, Codex CLI passes 57.6\% of the target-hidden instances, against 32.7\% for OpenHands and 9.7\% for Aider.
\stepcircle{2}~\textit{Reasoning effort}: for Codex/DS, high instead of no reasoning raises Pass@1 from 57.6\% to 82.8\% and KillComplete@1 from 23.6\% to 33.8\%.
\stepcircle{3}~\textit{LLM}: with Codex CLI at high effort, DS passes more instances (82.8\%) than the cheaper GPT-6 Luna (66.2\%) and the costliest LLM, GPT-6 Sol (73.5\%; Section~\ref{sec:rq2}).
Its KillComplete@1 is 5.5~pp [2.2, 8.2] above Luna's and 2.5~pp [$-$1.0, 5.4] above Sol's.
Under Claude Code, Claude Sonnet 5.5 reaches about the same Pass@1 and KillComplete@1 as DS (80.9\% and 80.5\%, 35.9\% and 34.4\%).
The leaderboard in our repository~\cite{req2testbenchwebsite} lists the scores of every configuration.

\noindent\textbf{Stronger configurations gain kill-complete tests by passing more instances, not by a higher kill-complete share of their passing tests.}
In the target-hidden view, Codex/DS has 336 more kill-complete tests than OpenHands/DS and 104 more than Codex/Luna, and the whole surplus comes from instances that only one of the two configurations passes.
On the instances that both pass, their tests are kill-complete about equally often (51.5\% against 54.4\% for OpenHands/DS, and 42.6\% against 44.0\% for Codex/Luna).

\par\medskip\noindent\fcolorbox{black!38}{black!5}{\parbox{\dimexpr\linewidth-2\fboxsep-2\fboxrule\relax}{
\small
\textbf{\sffamily Answer to RQ1.}
\enspace Agents often produce a passing test but far less often a behaviorally adequate one. 
The best-passing configuration, Codex/DS, passes 82.8\% of target-hidden instances but is kill-complete on only 33.8\%. 
The lead of Codex/DS over OpenHands/DS and Codex/Luna comes from passing more instances, not from a higher kill-complete share of its passing tests.}}
\par\medskip

\subsection{RQ2: Efficiency}\label{sec:rq2}

\begin{table}[t]
\centering
\caption{Cost and resource usage per generation in the target-hidden view.}
\label{tab:efficiency}
\footnotesize
\setlength{\tabcolsep}{3pt}
\begin{tabular}{@{}lcccccccc@{}}
\toprule
 & \textbf{Aider} & \textbf{OpenHands} & \multicolumn{2}{c}{\textbf{Claude Code}} & \multicolumn{4}{c}{\textbf{Codex CLI}} \\
\cmidrule(lr){4-5}\cmidrule(lr){6-9}
 & \textbf{DS, off} & \textbf{DS, off} & \textbf{DS, high} & \textbf{Sonnet, high} & \textbf{DS, off} & \textbf{DS, high} & \textbf{Luna, high} & \textbf{Sol, high} \\
\midrule
\$ per generation & 0.061 & 0.187 & 0.158 & 0.680 & 0.121 & 0.236 & 0.039 & 1.16 \\
\$ per kill-complete test & 1.56 & 1.21 & 0.49 & 1.99 & 0.55 & 0.74 & 0.15 & 3.92 \\
Time (minutes) & 2.0 & 5.7 & 6.2 & 2.7 & 4.1 & 8.7 & 3.2 & 3.1 \\
LLM calls & 2.0 & 137.8 & 88.9 & 22.3 & 96.6 & 102.4 & 23.6 & 16.8 \\
\bottomrule
\end{tabular}
\end{table}

\begin{figure}[t]
  \centering
  \includegraphics[width=\linewidth]{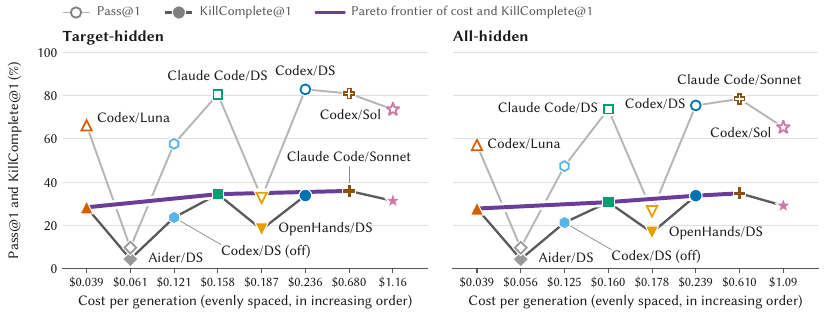}
  \caption{KillComplete@1 (filled) and Pass@1 (hollow) of the configurations, ordered by their cost per generation. 
  The thick line joins the Pareto frontier of the point estimates, i.e., the configurations whose KillComplete@1 no other reaches at the same or lower cost.}
  \label{fig:frontier}
  \vspace{-10pt}
\end{figure}

\noindent\textbf{Cost varies about 30-fold, and effectiveness does not follow it.}
As shown in \autoref{tab:efficiency}, a target-hidden generation costs from \$0.039 for Codex/Luna to \$1.16 for Codex/Sol on average, although a Codex/Sol generation takes only 3.1 minutes and 16.8 LLM calls.
\autoref{fig:frontier} orders the configurations by monetary cost: neither Pass@1 nor KillComplete@1 rises with it.
In the target-hidden view, only Codex/Luna, Claude Code/DS, and Claude Code/Sonnet lie on the Pareto frontier of cost and KillComplete@1.
Every other configuration costs more than one of them without a higher KillComplete@1.

\noindent\textbf{Spending more yields sharply diminishing returns.}
Moving along this frontier from Codex/Luna to Claude Code/DS multiplies the cost by 4.0 and raises KillComplete@1 by 6.1~pp [2.8, 8.8].
Beyond that, additional spending buys no measurable gain.
Codex/DS costs 1.5 times as much as Claude Code/DS for a KillComplete@1 that does not differ measurably ($-$0.6~pp [$-$2.4, 1.1]), and Codex/Sol 7.4 times as much for one that is no higher ($-$3.1~pp [$-$5.8, 0.1]).
With Codex CLI, moving from GPT-6 Luna to DS multiplies the cost by 6.0 for 5.5~pp, and moving on to GPT-6 Sol by another 4.9 for no further gain: Sol costs 30 times as much as its sibling Luna for 3.0~pp [1.1, 5.0].
Reasoning effort also raises the price of success.
It roughly doubles Codex/DS's cost per generation for 10.2~pp more KillComplete@1, and a kill-complete test costs \$0.74 instead of \$0.55.

\par\medskip\noindent\fcolorbox{black!38}{black!5}{\parbox{\dimexpr\linewidth-2\fboxsep-2\fboxrule\relax}{\small\textbf{\sffamily Answer to RQ2.}\enspace Performance of agents does not follow cost. Along the target-hidden Pareto frontier, 4.0 times the cost buys 6.1~pp of KillComplete@1, and more spending buys no measurable gain.}}\par\medskip

\subsection{RQ3: Associated Factors}\label{sec:rq3}

\noindent\textbf{Hiding the surrounding implementations costs passes more than kill-complete tests.}
As shown in \autoref{tab:overall-performance}, on the same instances, the all-hidden view lowers the Pass@1 of the five high-effort configurations by 2.5--9.1~pp (least for Claude Code/Sonnet), most of all on CoreShop (7.8--39.5~pp), but their KillComplete@1 by only 0.1--3.7~pp, which for Codex/DS, Codex/Luna, and Claude Code/Sonnet is too little to tell from no drop.
Reading the surrounding production code thus mainly helps a test run.
Without it, Codex/DS and Codex/Luna tests more often violate the runtime contracts of the interfaces that they call (see Section~\ref{sec:rq4}).

\noindent\textbf{Longer Scenarios and larger ground-truth tests are harder to pass and have more kills to reproduce.}
We rank each project's instances by a characteristic and compare the third with the lowest values to the third with the highest (five high-effort configurations, target-hidden view).
From the shortest to the longest Scenarios (a median of 5 against 15 Steps), Pass@1 falls from 82.8\% to 70.8\% and KillComplete@1 from 45.8\% to 22.7\%.
Larger ground-truth tests show the same pattern, by lines, covered production lines, or kills: Pass@1 falls by 6--14~pp and KillComplete@1 by 24--32~pp.
Repository-level characteristics matter less for kill-completeness.
With more hidden code or more target interfaces, both metrics fall by 11--15~pp.

\par\medskip\noindent\fcolorbox{black!38}{black!5}{\parbox{\dimexpr\linewidth-2\fboxsep-2\fboxrule\relax}{
\small
\textbf{\sffamily Answer to RQ3.}
\enspace 
The all-hidden view lowers Pass@1 by 2.5--9.1~pp but KillComplete@1 by only 0.1--3.7~pp. 
Longer Scenarios and larger ground-truth tests are harder to pass and have more kills to reproduce.
}
}
\par\medskip

\subsection{RQ4: Failure Modes}\label{sec:rq4}

\begin{figure}[t]
  \centering
  \includegraphics[width=\linewidth]{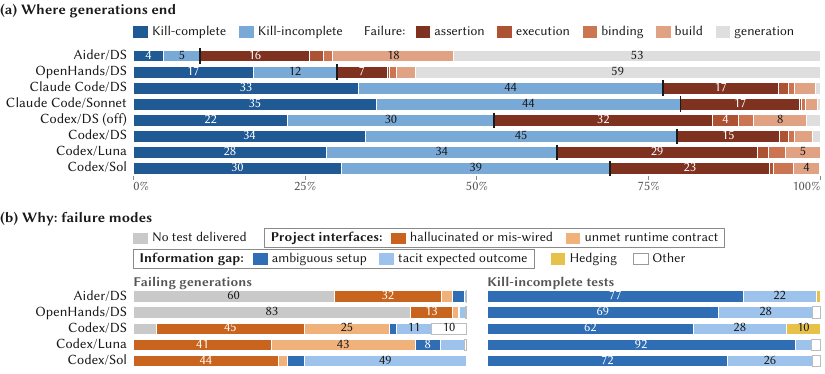}
  \caption{
  (a)~Where generations end, in percent of each configuration's generations (project means over both views).
  Black ticks mark the end of the passing tests. 
  (b)~Why: failure modes of failing generations and kill-incomplete tests, in percent.}
  \label{fig:outcomes}
  \vspace{-12pt}
\end{figure}

We analyze both failing generations, which produce no passing tests, and passing but kill-incomplete tests.
For each of five configurations, we drew a stratified random sample of 40 cases of each type, yielding 200 failing generations and 200 kill-incomplete tests.
We derived the failure modes in \autoref{fig:outcomes}(b) through a bottom-up analysis of the sampled cases' outcomes and code, supplemented by missed-mutant analysis for kill-incomplete tests.
We assigned labels using deterministic rules and an LLM judge (Claude Opus 5.5), with manual checks by the authors.

\noindent\textbf{Failing generations mostly deliver no test or misuse the project's interfaces.}
\stepcircle{1}~\textit{No test delivered}: 
As shown in \autoref{fig:outcomes}(a) (grey bar), Aider/DS and OpenHands/DS deliver no Test Implementation in 53.4\% and 59.0\% of their generations.
Aider/DS's reply mostly repeats itself up to its output limit, and OpenHands/DS mostly exhausts its 200 LLM calls on exploration, without writing a test file.
As shown in \autoref{tab:efficiency}, with the same call limit and LLM, Codex CLI and Claude Code deliver a test with about 100 calls.
\stepcircle{2}~\textit{Hallucinated or miswired interfaces} (45\% and 41\% of the failing generations of Codex/DS and Codex/Luna, \autoref{fig:outcomes}(b), left): 
The test uses classes, methods, signatures, or services that do not exist in the project, such as \texttt{Tools::link\_rewrite()} in 40 Codex/Luna generated tests on PrestaShop, redefines existing Steps, or misuses the BDD framework.
\stepcircle{3}~\textit{Unmet runtime contracts} (25\% and 43\%): 
The test calls existing interfaces in ways that their implementation rejects, e.g., an operation that the concrete class does not support.
Declarations do not reveal such contracts, whereas the surrounding implementations often do.
The mode is thus more frequent when all implementations are hidden (34\% and 51\%, against 13\% and 32\% in the target-hidden view).
Another 13\% and 15\% fail on the information gap below.

\noindent\textbf{Kill-incomplete tests fall short where the Requirement leaves information unstated.}
This information gap accounts for 90--99\% of each judged configuration's kill-incomplete tests (\autoref{fig:outcomes}(b), right).
In 120 of the 200 judged tests, the missing information is in the agent's own workspace, e.g., in fixture files or helpers.
In most of the remaining 80, the needed project convention of setup or check appears only in developers' tests, which the benchmark removes from the agent's workspace.
\stepcircle{1}~\textit{Ambiguous setup} (over 50\%): the Step names a state that can be established in several plausible ways, and the test takes a shorter but unexpected way than the ground truth, as in \autoref{fig:example}.
For example, on python-docx instances whose ground truth opens a fixture file, tests that also open one are kill-complete in 81.8\% of cases, against 24.6\% for tests that construct the state themselves.
\stepcircle{2}~\textit{Tacit expected outcome} (up to 28\%): 
the \textit{Then} Step states only part of what the ground truth checks, and what else to check is tacit for developers with domain knowledge.
For example, on 25 Algorand SDK Scenarios, the Steps name a few fields of a response, the ground truth compares the whole response, and only 9 of the 320 passing tests are kill-complete, all of them by Claude Code/Sonnet.
Ambiguous setup and tacit expected outcomes are normal in requirements: a requirement describes behavior in terms of the domain~\cite{zave1997four}, and the knowledge that completes it stays tacit with the developers~\cite{ferrari2016ambiguity}.
An agent must recover this knowledge from the project, or the requirement must be refined to state it.
\stepcircle{3}~\textit{Hedging} (10\% for Codex/DS): fallbacks or broad exception handlers let Steps pass whatever the product does.

\par\medskip\noindent\fcolorbox{black!38}{black!5}{\parbox{\dimexpr\linewidth-2\fboxsep-2\fboxrule\relax}{\small\textbf{\sffamily Answer to RQ4.}\enspace Failing generations mostly deliver no test or misuse the project's interfaces. 
Kill-incomplete tests fall short where the setup is ambiguous or the expected outcome tacit (90--99\%).}}\par\medskip

\subsection{Toward Requirement-Grounded Harnesses}\label{sec:probe}

Four preliminary studies investigate what a harness around the agent can supply.

\noindent\textbf{Grounding tests in project knowledge} targets ambiguous setup.
An added prompt paragraph instructs the agent to establish each \textit{Given} through the project's own fixtures, helpers, and entry points, not by constructing objects.
On 231 python-docx and Algorand SDK instances not used to design the paragraph, this raises KillComplete@1 by 6.1~pp, 15.0~pp, and 8.0~pp for Codex/DS, Codex/Luna, and Codex/Sol ($p\leq0.005$).

\noindent\textbf{Enhancing the Requirement} targets ambiguous setup and tacit expected outcomes.
In domain language and naming no code, one author modifies the prompt to clarify what the ground-truth test sets up or checks beyond what a Step states.
For example, the unchanged Step ``the parsed Get Block response should have rewards pool \dots'' is clarified as ``the parsed response matches the mock http response as a whole''.
As a result, Codex/Luna's kill-complete tests rise from 2 to 14 on 17 Algorand SDK Scenarios with tacit outcomes ($p<0.001$), and from 25 to 36 on 60 python-docx instances with an ambiguous \textit{Given}.
Enhancing requirements automatically with the project's conventions, fixtures, and helpers is a potential direction.

\noindent\textbf{Decomposing the Requirement} targets misused project interfaces.
Longer Scenarios are harder to pass (Section~\ref{sec:rq3}), but many of their Steps are shared with smaller Scenarios that they depend on.
A harness could therefore decompose the target requirement and generate tests for the resulting smaller requirements first, and give them to the agent for the test generation of target requirement.
We repeated test generation for 100 requirements from CoreShop and Kapua for which Codex/Luna had initially failed to produce passing tests.
Retries with the unchanged prompt produced passing tests for 39 requirements due to the agent's non-determinism.
Augmenting the prompt with Codex/Luna's own previously generated passing tests for smaller \texttt{depends\_on} requirements increased this number to 74 ($p<0.001$), suggesting that reusing these tests improves success beyond retries alone.

\noindent\textbf{Exploiting non-determinism.} 
Non-determinism makes a single generation unreliable but leaves room for improvement.
We observed that the best of three generations, chosen by the evaluator, is kill-complete 8.6--10.2~pp more often than a single one (e.g., for Codex/DS, 44.8\% against 36.2\% on the repeated sample).
A harness can exploit this only if it recognizes the behaviorally adequate candidate.

\section{Threats to Validity}

\noindent\textbf{Construct validity.}
KillComplete@1 measures behavioral adequacy against one developer-written test, so a kill-incomplete test may still validly test a narrower reading of the requirement.
However, the gap persists when only check-dependent kills count, and most kill-incomplete tests run a faulty line and still pass (Section~\ref{sec:rq1}), so the gap does not stem only from the test's setup route.
Moreover, mutants are artificial faults, but their detection correlates with that of real faults~\cite{just2014mutants}.
Equivalent mutants, which no test can kill, do not lower any score, because we count only the mutants that the ground-truth test kills.

\noindent\textbf{Internal validity.}
Due to the time budgets, the experiments did not repeat all agents' test generations multiple times on the whole benchmark to reduce the effects of agents' non-determinism.
We choose five configurations (Codex/DS, Codex/Luna, Claude Code/DS, OpenHands/DS, and Aider/DS) to regenerate twice to assess the effects.
Up to 24\% of their pass verdicts and 17\% of their kill-complete verdicts flip, but their Pass@1 and KillComplete@1 change by at most 4.5~pp.
Additionally, the target-hidden view rests on an LLM-drafted, author-accepted map of the target implementation, which may hide slightly more or less than the code that realizes the target behavior.
However, the all-hidden view does not depend on this map, and it shows the same gap between passing and kill-complete tests.
Finally, the projects are public and may be in the training data of some LLMs.
But memorized developer tests would raise KillComplete@1 and thus narrow, not widen, the gap that we report.

\noindent\textbf{External validity.}
The eight projects were selected purposively, but they cover five languages, four BDD frameworks, six domains, and three test levels, and our reported gap appears in every one of them.
Moreover, Gherkin Scenarios are more structured than most informal requirements.
Informal requirements are likely to state even less of the setup and the expected outcome, so we expect the information gap to be larger for them, not smaller.
Finally, both views in the benchmark remove the tests of all Scenarios, so that none reveals part of the ground truth.
However, in practice, a developer who extends a test suite could read them.
Our results thus concern writing a test without the tests of other Scenarios.
We leave such cases in future work.

\section{Conclusion}
We presented \tool, a benchmark of 5,086 execution-validated requirement-to-test instances from 8 real projects. It evaluates a generated test by whether it passes and reproduces the fault detection of the developers' own test.
We evaluated eight agent--model--reasoning configurations, and found that the best-passing configuration passes 82.8\% of the instances but is kill-complete on only 33.8\%.
Kill-incomplete tests mostly fall short where the requirement leaves the setup ambiguous or the expected outcome tacit. They take shortcuts instead of the project's own fixtures and entry points, or check less than the developers' tests.
This gap persists even with costlier agent configurations, motivating requirement-grounded harnesses for test generation that guide requirement understanding, test validation, and project-specific knowledge acquisition.

\section{Data Availability}
The benchmark code and evaluation results are available at \url{https://github.com/code-philia/Req2Test-Bench}, and the dataset is available at \url{https://huggingface.co/datasets/code-philia/Req2Test-Bench}.

\bibliographystyle{ACM-Reference-Format}
\bibliography{sample-base}

@article{qi2026generatingprojectspecifictestcases,
author = {Qi, Binhang and Lin, Yun and Weng, Xinyi and Huang, Yuhuan and Liu, Chenyan and Sun, Hailong and Jin, Zhi and Dong, Jin Song},
title = {Generating Project-Specific Test Cases with Requirement Validation Intention},
year = {2026},
issue_date = {October 2026},
publisher = {Association for Computing Machinery},
address = {New York, NY, USA},
volume = {3},
number = {ISSTA},
url = {https://doi.org/10.1145/3832181},
doi = {10.1145/3832181},
journal = {Proc. ACM Softw. Eng.},
month = oct,
articleno = {ISSTA090},
numpages = {23}
}

@INPROCEEDINGS{IntUT,
  author={Nan, Zifan and Guo, Zhaoqiang and Liu, Kui and Xia, Xin},
  booktitle={2025 IEEE/ACM 47th International Conference on Software Engineering (ICSE)}, 
  title={Test Intention Guided LLM-Based Unit Test Generation}, 
  year={2025},
  volume={},
  number={},
  pages={1026-1038}
}

@inproceedings{liu2026e2edev,
  title={E2EDev: Benchmarking Large Language Models in End-to-End Software Development Task},
  author={Liu, Jingyao and Huang, Chen and Guan, Zhizhao and Lei, Wenqiang and Deng, Yang},
  booktitle={Proceedings of the 64th Annual Meeting of the Association for Computational Linguistics (Volume 1: Long Papers)},
  pages={35032--35068},
  year={2026}
}

@misc{claudesonnet55,
  author       = {{Anthropic}},
  title        = {Claude Sonnet 5.5},
  year         = {2026},
  howpublished = {\url{https://www.anthropic.com/claude-sonnet-5-5}}
}

@inproceedings{junior2026givenwhenthen,
  title={GivenWhenThen: A Dataset of BDD Test Scenarios Mined from Open Source Projects},
  author={J{\'u}nior, Luciano Belo de Alc{\^a}ntara and Montandon, Jo{\~a}o Eduardo},
  booktitle={Proceedings of the 23rd International Conference on Mining Software Repositories},
  pages={595--599},
  year={2026}
}

@inproceedings{francca2026req2test,
  title={Req2test: an approach for the automatic generation of unit tests from user stories using llms},
  author={Fran{\c{c}}a, Mattheus and Benitti, Fabiane},
  booktitle={Proceedings of the 41st ACM/SIGAPP Symposium on Applied Computing},
  pages={1654--1663},
  year={2026}
}

@misc{li2024devbench,
  title         = {{DevBench}: A Comprehensive Benchmark for Software Development},
  author        = {Bowen Li and Wenhan Wu and Ziwei Tang and Lin Shi and John Yang and Jinyang Li and Shunyu Yao and Chen Qian and Binyuan Hui and Qicheng Zhang and Zhiyin Yu and He Du and Ping Yang and Dahua Lin and Chao Peng and Kai Chen},
  year          = {2024},
  eprint        = {2403.08604v1},
  archivePrefix = {arXiv},
  primaryClass  = {cs.CL},
  url           = {https://arxiv.org/abs/2403.08604v1}
}

@misc{yang2026uncoveringbusinesslogicbugs,
      title={Uncovering Business Logic Bugs via Semantics-Driven Unit Test Generation}, 
      author={Chen Yang and Junjie Chen},
      year={2026},
      eprint={2604.23509},
      archivePrefix={arXiv},
      primaryClass={cs.SE},
      url={https://arxiv.org/abs/2604.23509}, 
}

@misc{codex,
  author       = {{OpenAI}},
  title        = {Lightweight coding agent that runs in your terminal},
  year         = {2026},
  howpublished = {\url{https://github.com/openai/codex}},
}

@misc{aider,
  author       = {{Aider}},
  title        = {aider is AI pair programming in your terminal},
  year         = {2026},
  howpublished = {\url{https://github.com/Aider-AI/aider}},
}

@misc{gpt6,
  author       = {{OpenAI}},
  title        = {Introducing GPT‑6 Sol and Luna},
  year         = {2026},
  howpublished = {\url{https://openai.com/index/introducing-gpt-6-sol-and-luna/}},
}

@misc{bubblewrap,
  author       = {{Bubblewrap Contributors}},
  title        = {Low-level unprivileged sandboxing tool used by Flatpak and similar projects},
  year         = {2026},
  url          = {https://github.com/containers/bubblewrap},
}

@misc{claude-code,
  author       = {claude-code Authors},
  title        = {Claude Code},
  howpublished = {\url{https://github.com/anthropics/claude-code}},
}

@inproceedings{wang2025openhandsopenplatformai,
title={OpenHands: An Open Platform for {AI} Software Developers as Generalist Agents},
author={Xingyao Wang and Boxuan Li and Yufan Song and Frank F. Xu and Xiangru Tang and Mingchen Zhuge and Jiayi Pan and Yueqi Song and Bowen Li and Jaskirat Singh and Hoang H. Tran and Fuqiang Li and Ren Ma and Mingzhang Zheng and Bill Qian and Daniel Shao and Niklas Muennighoff and Yizhe Zhang and Binyuan Hui and Junyang Lin and Robert Brennan and Hao Peng and Heng Ji and Graham Neubig},
booktitle={The Thirteenth International Conference on Learning Representations},
year={2025},
url={https://openreview.net/forum?id=OJd3ayDDoF}
}

@inproceedings{tufano2022methods2test,
  title={Methods2Test: A dataset of focal methods mapped to test cases},
  author={Tufano, Michele and Deng, Shao Kun and Sundaresan, Neel and Svyatkovskiy, Alexey},
  booktitle={Proceedings of the 19th International Conference on Mining Software Repositories},
  pages={299--303},
  year={2022}
}

@article{zhang2024testbench,
  title={Testbench: Evaluating class-level test case generation capability of large language models},
  author={Zhang, Quanjun and Shang, Ye and Fang, Chunrong and Gu, Siqi and Zhou, Jianyi and Chen, Zhenyu},
  journal={arXiv preprint arXiv:2409.17561},
  year={2024}
}

@article{mundler2024swt,
  title={Swt-bench: Testing and validating real-world bug-fixes with code agents},
  author={M{\"u}ndler, Niels and M{\"u}ller, Mark N and He, Jingxuan and Vechev, Martin},
  journal={Advances in Neural Information Processing Systems},
  volume={37},
  pages={81857--81887},
  year={2024}
}

@inproceedings{alian2025feature,
  title={Feature-driven end-to-end test generation},
  author={Alian, Parsa and Nashid, Noor and Shahbandeh, Mobina and Shabani, Taha and Mesbah, Ali},
  booktitle={2025 IEEE/ACM 47th International Conference on Software Engineering (ICSE)},
  pages={450--462},
  year={2025},
  organization={IEEE}
}

@misc{cucumberGherkin,
  author = {{The Cucumber Open Source Project}},
  title  = {Gherkin Reference},
  year   = {2026},
  month  = aug,
  url    = {https://cucumber.io/docs/gherkin/reference/},
}

@inproceedings{jain2025testgeneval,
  title     = {{TestGenEval}: A Real World Unit Test Generation and Test Completion Benchmark},
  author    = {Kush Jain and Gabriel Synnaeve and Baptiste Rozi{\`e}re},
  booktitle = {The Thirteenth International Conference on Learning Representations},
  year      = {2025},
  url       = {https://openreview.net/forum?id=7o6SG5gVev}
}

@inproceedings{wang2015umtg,
  title={UMTG: a toolset to automatically generate system test cases from use case specifications},
  author={Wang, Chunhui and Pastore, Fabrizio and Goknil, Arda and Briand, Lionel C and Iqbal, Zohaib},
  booktitle={Proceedings of the 2015 10th Joint Meeting on Foundations of Software Engineering},
  pages={942--945},
  year={2015}
}

@inproceedings{fischbach2020specmate,
  title={Specmate: Automated creation of test cases from acceptance criteria},
  author={Fischbach, Jannik and Vogelsang, Andreas and Spies, Dominik and Wehrle, Andreas and Junker, Maximilian and Freudenstein, Dietmar},
  booktitle={2020 IEEE 13th international conference on software testing, validation and verification (ICST)},
  pages={321--331},
  year={2020},
  organization={IEEE}
}

@article{fischbach2021cira,
  title={Cira: A tool for the automatic detection of causal relationships in requirements artifacts},
  author={Fischbach, Jannik and Frattini, Julian and Vogelsang, Andreas},
  journal={arXiv preprint arXiv:2103.06768},
  year={2021}
}

@inproceedings{
    jimenez2024swebench,
    title={{SWE}-bench: Can Language Models Resolve Real-world Github Issues?},
    author={Carlos E Jimenez and John Yang and Alexander Wettig and Shunyu Yao and Kexin Pei and Ofir Press and Karthik R Narasimhan},
    booktitle={The Twelfth International Conference on Learning Representations},
    year={2024},
    url={https://openreview.net/forum?id=VTF8yNQM66}
}

@article{chen2022codet,
  title={Codet: Code generation with generated tests},
  author={Chen, Bei and Zhang, Fengji and Nguyen, Anh and Zan, Daoguang and Lin, Zeqi and Lou, Jian-Guang and Chen, Weizhu},
  journal={arXiv preprint arXiv:2207.10397},
  year={2022}
}

@article{huang2023agentcoder,
  title={Agentcoder: Multi-agent-based code generation with iterative testing and optimisation},
  author={Huang, Dong and Zhang, Jie M and Luck, Michael and Bu, Qingwen and Qing, Yuhao and Cui, Heming},
  journal={arXiv preprint arXiv:2312.13010},
  year={2023}
}

@article{schafer2023empirical,
author = {Sch{"a}fer, Max and Nadi, Sarah and Eghbali, Aryaz and Tip, Frank},
title = {An Empirical Evaluation of Using Large Language Models for Automated Unit Test Generation},
year = {2024},
issue_date = {Jan. 2024},
publisher = {IEEE Press},
volume = {50},
number = {1},
issn = {0098-5589},
url = {https://doi.org/10.1109/TSE.2023.3334955},
doi = {10.1109/TSE.2023.3334955},
journal = {IEEE Trans. Softw. Eng.},
month = jan,
pages = {85––105},
numpages = {21}
}

@article{hu2026difftestgen,
  title={DiffTestGen: Change-Directed LLM-Based Testing for Exposing Behavioral Differences},
  author={Hu, Huimin and Cadar, Cristian and Pradel, Michael},
  journal={arXiv preprint arXiv:2607.16024},
  year={2026}
}

@article{ahmed2024tddbench,
  title={Tdd-bench verified: Can llms generate tests for issues before they get resolved?},
  author={Ahmed, Toufique and Hirzel, Martin and Pan, Rangeet and Shinnar, Avraham and Sinha, Saurabh},
  journal={arXiv preprint arXiv:2412.02883},
  year={2024}
}

@misc{shi2026autoglue,
      title={Bridging Behavior and Implementation: Automated Java Glue Code Generation for Behavior-Driven Development}, 
      author={Xinyu Shi and Zhou Yang and An Ran Chen},
      year={2026},
      eprint={2607.19703},
      archivePrefix={arXiv},
      primaryClass={cs.SE},
      url={https://arxiv.org/abs/2607.19703}, 
}

@inproceedings{ferreira2025acceptance,
  title={Acceptance test generation with large language models: An industrial case study},
  author={Ferreira, Margarida and Viegas, Luis and Faria, Joao Pascoal and Lima, Bruno},
  booktitle={2025 IEEE/ACM International Conference on Automation of Software Test (AST)},
  pages={1--11},
  year={2025},
  organization={IEEE}
}

@INPROCEEDINGS{fonseca2025atomic,
  author={Fonseca, Pedro Luís and Lima, Bruno and Faria, João Pascoal},
  booktitle={2025 40th IEEE/ACM International Conference on Automated Software Engineering (ASE)}, 
  title={Streamlining Acceptance Test Generation for Mobile Applications Through Large Language Models: An Industrial Case Study}, 
  year={2025},
  volume={},
  number={},
  pages={3311-3321},
  doi={10.1109/ASE63991.2025.00273}}

@article{jia2011mutation,
author = {Jia, Yue and Harman, Mark},
title = {An Analysis and Survey of the Development of Mutation Testing},
year = {2011},
issue_date = {September 2011},
publisher = {IEEE Press},
volume = {37},
number = {5},
issn = {0098-5589},
url = {https://doi.org/10.1109/TSE.2010.62},
doi = {10.1109/TSE.2010.62},
journal = {IEEE Trans. Softw. Eng.},
month = sep,
pages = {649––678},
numpages = {30}
}

@inproceedings{just2014mutants,
  title={Are mutants a valid substitute for real faults in software testing?},
  author={Just, Ren{\'e} and Jalali, Darioush and Inozemtseva, Laura and Ernst, Michael D and Holmes, Reid and Fraser, Gordon},
  booktitle={Proceedings of the 22nd ACM SIGSOFT international symposium on foundations of software engineering},
  pages={654--665},
  year={2014}
}

@misc{pit_mutators,
  author       = {{PIT}},
  title        = {Java Mutation Operators},
  howpublished = {\url{https://pitest.org/quickstart/mutators/}},
}

@misc{infection_mutators,
  author       = {{Infection}},
  title        = {PHP Mutators},
  howpublished = {\url{https://infection.github.io/guide/mutators.html}},
}

@article{zave1997four,
author = {Zave, Pamela and Jackson, Michael},
title = {Four dark corners of requirements engineering},
year = {1997},
issue_date = {Jan. 1997},
publisher = {Association for Computing Machinery},
address = {New York, NY, USA},
volume = {6},
number = {1},
issn = {1049-331X},
doi = {10.1145/237432.237434},
journal = {ACM Trans. Softw. Eng. Methodol.},
month = jan,
pages = {1–30},
numpages = {30},
}

@article{ferrari2016ambiguity,
author = {Ferrari, Alessio and Spoletini, Paola and Gnesi, Stefania},
title = {Ambiguity and tacit knowledge in requirements elicitation interviews},
year = {2016},
issue_date = {September 2016},
publisher = {Springer-Verlag},
address = {Berlin, Heidelberg},
volume = {21},
number = {3},
issn = {0947-3602},
url = {https://doi.org/10.1007/s00766-016-0249-3},
doi = {10.1007/s00766-016-0249-3},
journal = {Requir. Eng.},
month = sep,
pages = {333–355},
numpages = {23},
}

@article{kong2026arc,
author = {Kong, Weiyu and Lin, Yun and Teoh, Xiwen and Nguyen, Duc-Minh and Ren, Ruofei and Chang, Jiaxin and Hu, Haoxu and Chen, Haoyu},
title = {Compiling Large Multi-modal Requirement Documents into Runnable Software Systems: From an Agentic Test-Driven Perspective},
year = {2026},
issue_date = {October 2026},
publisher = {Association for Computing Machinery},
address = {New York, NY, USA},
volume = {3},
number = {ISSTA},
url = {https://doi.org/10.1145/3832196},
doi = {10.1145/3832196},
journal = {Proc. ACM Softw. Eng.},
month = oct,
articleno = {ISSTA105},
numpages = {24}
}

@misc{req2testbenchwebsite,
  author       = {{Req2Test-Bench Authors}},
  title        = {Req2Test-Bench},
  year         = {2026},
  url          = {https://github.com/code-philia/Req2Test-Bench},
}

@misc{deepseekai2026deepseekv41flashpushinglimitskv,
      title={DeepSeek-V4.1-Flash: Pushing the Limits of KV Cache Compression}, 
      author={{DeepSeek-AI}},
      year={2026},
      eprint={2609.19969},
      archivePrefix={arXiv},
      primaryClass={cs.CL},
      url={https://arxiv.org/abs/2609.19969}, 
}

\end{document}